\documentclass[journal]{IEEEtranTIE}
\usepackage{graphicx}
\usepackage{subfigure}  
\usepackage{float}
\usepackage{cite}
\usepackage{fancyhdr}
\usepackage{picinpar}
\usepackage{amsmath,amsfonts}
\usepackage{url}
\usepackage{flushend}
\usepackage[latin1]{inputenc}
\usepackage{colortbl}
\usepackage{soul}
\usepackage{multirow}
\usepackage{pifont}
\usepackage{color}
\usepackage{alltt}
\usepackage{enumerate}
\usepackage{siunitx}
\usepackage{breakurl}
\usepackage{epstopdf}
\usepackage{pbox}
\usepackage{mathrsfs}
\usepackage{algorithm}
\usepackage{algorithmicx}  
\usepackage{algpseudocode}  
\usepackage{color}
\usepackage{booktabs}
\usepackage[colorlinks,linkcolor=red, anchorcolor=blue,citecolor=blue]{hyperref}
\definecolor{gry}{rgb}{0.6,0.6,0.6} 
\definecolor{red}{rgb}{1,0,0}

\newtheorem{assumption}{\textbf{Assumption}}
\newtheorem{remark}{\textbf{Remark}}
\newtheorem{thm}{\textbf{Theorem}}
\newtheorem{lemma}{\textbf{Lemma}}

\newtheorem{definition}{\textbf{Definition}}

\begin{document}
\title{An Improved Level Set Method for Reachability Problems in Differential Games \\ 
\normalsize{\textcolor{red}{This work has been submitted to the IEEE for possible publication. Copyright may be transferred without notice, after which this version may no longer be accessible.}}}

\author{
	\vskip 1em
	
	Wei Liao, \emph{Member, IEEE},
	Taotao Liang,
	Pengwen Xiong, \emph{Member, IEEE},
	Chen Wang, Aiguo Song, \emph{Senior Member, IEEE},
	and Peter X. Liu, \emph{Fellow, IEEE}

	\thanks{
	
		Manuscript received Month xx, 2xxx; revised Month xx, xxxx; accepted Month x, xxxx.
		This work was supported in part by the xxx Department of xxx under Grant  (sponsor and financial support acknowledgment goes here).
		
		(Authors' names and affiliation) First A. Author1 and Second B. Author2 are with the xxx Department, University of xxx, City, Zip code, Country, on leave from the National Institute for xxx, City, Zip code, Country (e-mail: author@domain.com). 
		
		Third C. Author3 is with the National Institute of xxx, City, Zip code, Country (corresponding author to provide phone: xxx-xxx-xxxx; fax: xxx-xxx-xxxx; e-mail: author@ domain.gov).
		Corresponding author: Pengwen Xiong.
	}
}

\maketitle
	
\begin{abstract}
	This study focuses on reachability problems in differential games. An improved level set method for computing reachable tubes is proposed in this paper.
	The reachable tube is described as a sublevel set of a value function, which is the viscosity solution of a Hamilton-Jacobi equation with running cost. We generalize the concept of reachable tubes and propose a new class of reachable tubes, which are referred to as cost-limited one. In particular, a performance index can be specified for the system, and a cost-limited reachable tube is a set of initial states of the system's trajectories that can reach the target set before the performance index increases to a given admissible cost.	Such a reachable tube can be obtained by specifying the corresponding running cost function for the Hamilton-Jacobi equation. Different non-zero sublevel sets of the viscosity solution of the Hamilton-Jacobi equation at a certain time point can be used to characterize the cost-limited reachable tubes with different admissible costs (or the reachable tubes with different time horizons), thus reducing the storage space consumption. Several examples are provided to illustrate the validity and accuracy of the proposed method.
\end{abstract}


% \markboth{IEEE TRANSACTIONS ON INDUSTRIAL ELECTRONICS}%
{}

\definecolor{limegreen}{rgb}{0.2, 0.8, 0.2}
\definecolor{forestgreen}{rgb}{0.13, 0.55, 0.13}
\definecolor{greenhtml}{rgb}{0.0, 0.5, 0.0}

\section{Introduction}
\IEEEPARstart{D}{ifferential} 
games, a classical category of problems in control theory, 
involve the modeling and analysis of conflicts among multiple players in a dynamic system.
It is generally accepted that the concept of differential games was first introduced by Isaacs \cite{anew1}.
It has been widely applied in several fields, such as engineering \cite{newa2,anew3,ieeecaa} and economics \cite{newa4}.

In recent years, some studies have introduced reachability analyses in differential game problems \cite{a008,a0015,newa5}.
In the context of differential games, reachability analysis typically involves two players, 
where one player aims to drive the system state away from the target set, 
while the other aims to drive the system state toward the target set. 
The goal of reachability analysis is to find a reachable tube (RT), i.e., 
a set of initial states of the system's trajectories that can reach the target set within a given time horizon \cite{a001,a008}.
By applying reachability analysis, a variety of engineering problems can be solved, especially
those involving system safety \cite{a008,newi1,ieeecaa2}.

However, computing RT is challenging. Except for some solvable problems \cite{newa6}, the exact RT computation is generally intractable.
Therefore, a common practice is to compute RT's approximate expressions.
Numerous techniques were proposed in the past decades.
They can be divided into the following categories: ellipsoidal methods \cite{i3,a006,i4}, polyhedral methods \cite{i2,a005,a007}, 
and methods based on state-space discretization.
The first two categories are able to solve high-dimensional problems but have strict requirements on the form of dynamic 
systems and are primarily used to solve linear problems.

The main advantage of state-space discretization-based approaches is that they have less stringent requirements for the 
form of a dynamic system; and they can be used to solve nonlinear problems.
This category of methods mainly includes the level set method \cite{a001,a002,a008} and distance
fields over grids (DFOG) method \cite{a0010,a0011,a0012}. The former is more widely used in engineering than the latter.

In the level set method, the Hamilton-Jacobi (HJ) partial differential equation (PDE) without running cost function is numerically solved, 
and the RT can then be expressed as the zero sublevel set of the solution of HJ PDE \cite{a001}.
Based on this principle, some toolboxes have been developed \cite{tr1,web1,web2,web3} and greatly promoted the 
development of computational techniques for RT.
Furthermore, they have been employed to solve many practical engineering problems, 
such as flight control systems \cite{i5,a009,a0013}, ground traffic management systems \cite{i6,a0014}, and air traffic management systems \cite{a0015}.

Although the level set method has developed into a mature, reliable, and commonly used tool for years, 
it has some shortcomings. It poses very high storage space requirements. 
To save RT with a given time horizon, one needs to save the solution of the HJ equation at a certain time point; 
the storage space grows significantly with an increase in the dimension of the problem. To save RTs with different time horizons, 
one needs to save the solutions of the 
HJ equation at different time points, which in turn leads to a significant increase in storage space consumption.

In addition, since its inception, differential games have been inextricably linked to optimal control problems. 
In both optimal control and differential game problems, different forms of performance indexes can be specified, 
and the time consumption is only a special form. A general form of the performance index typically contains a Lagrangian \cite{cyb008}, 
which is the time integral of the running cost. 
Therefore, it is necessary to develop a novel framework that can be used to determine whether a target set can be reached within an admissible performance index. 
At the same time, such a framework must have great engineering significance. For example, fuel consumption is a common performance index in some engineering fields, 
and researchers may be concerned about whether a target set can be reached before running out of fuel.
Although it is theoretically possible to transform this new class of problems into classical reachability problems 
by considering the performance index as an extended system state, which can then be solved using level set method, 
adding one dimension to the state space increases 
the computation time and memory consumption by a factor of tens or even hundreds (depending on the number of grids in this added dimension).

To overcome the shortcomings of the level set method and determine reachability under different forms of running cost, 
we propose an improved level set method to solve reachability problems and a more generalized definition of RT. 
We thus aim to make the following contributions:

\begin{itemize}
    \item [(1)] We propose an improved level set method for computing the RT. 
	In the proposed method, RT is represented as  
	a non-zero sublevel set of a value function, which is the viscosity solution of an HJ PDE with constant 1 as a running cost function. 
	The RTs with different time horizons can 
	be represented as a family of non-zero sublevel sets of this value function,
	and thus reduce the consumption of storage space.

	\item [(2)] We define a cost-limited reachable tube (CRT). A running cost function can be specified for the system. CRT is a set of states that can be driven 
	into the target set before the performance index increases to the given admissible cost.
	
	\item [(3)] We characterize CRTs into a family of non-zero sublevel sets of a value function
    obtained by solving an HJ PDE with a corresponding running cost function.
	
\end{itemize}

The remainder of this paper is structured as follows. Section II introduces RT and the level set method. 
Section III describes the proposed method in detail. Section IV
generalizes the
reachability problem and presents the definition of CRT, and also 
presents how to compute the CRT by using the proposed method.
Numerical examples are provided in Section V to illustrate the validity of the proposed method. 
Section VI summarizes the results of this study.

\section{Reachable tube and the level set method}
Consider a continuous time control system with a fully observable state:  
\begin{align}
	\label{sys1}
    \dot{s}=f(s,a,b)
\end{align}
where $s\in \mathbb{R}^n$ is system state, $a\in \mathcal{A}$ is the control input of Player I,
and $b\in \mathcal{B}$ is the control input of Player II.
The function $f(.,.,.):\mathbb{R}^n\times \mathcal{A} \times \mathcal{B} \to \mathbb{R}^n$ is 
Lipschitz continuous and bounded. Let $\mathscr{A}_t$ and $\mathscr{B}_t$ denote the set of Lebesgue 
measurable functions from the interval $[t,\infty)$ to $\mathcal{A}$ and $\mathcal{B}$, respectively.
Then, given the initial time $t$ and state $s$, under control inputs $a(.)\in \mathscr{A}_t$ and $b(.) \in \mathscr{B}_t$, 
the evolution of system (\ref{sys1}) is determined by a continuous trajectory
$\phi_{s,t}^{a,b}(.):[t,\infty)\to \mathbb{R}^n$ and satisfies:
\begin{align}
    \begin{split}
        &\frac{d}{d\tau} \phi_{s,t}^{a,b}(\tau)=f\left(\phi_{s,t}^{a,b}(\tau),a(\tau),b(\tau)\right) \forall \tau \in [t,\infty)\\
        &\phi_{s,t}^{a,b}(t)=s
    \end{split}
\end{align}
We denote the target set as $K$, and assume that Player I aims to drive the system state toward the target set, 
and Player II aims to drive the system state away from it.

In a differential game, different information patterns provide advantages to different players;
in particular, players who play nonanticipative strategies have an advantage over their opponents \cite{a008}.
In this study, because the target set represents a safe portion of the state space, 
we prefer to underapproximate the RTs (or CRTs) of the target set rather than overapproximate them. 
To consider the worst case, we allow Player II, the player trying to drive the system state away from the target set, 
to use nonanticipative strategies. 
The set containing all the nonanticipative strategies for Player II is denoted as:

\begin{align}
    \begin{split}
        \mathfrak{B}_t=&\bigg\{\beta: \mathscr{A}_t\to \mathscr{B}_t \big| 
         \forall r \in [t,\infty), \forall a(.),\hat{a}(.)\in \mathscr{A}_t, \\
        &\big( a(\tau)=\hat{a}(\tau) \text{ a.e. }\tau\in [t,r] \Longrightarrow   \\
         & \beta[a](\tau)=\beta[\hat{a}](\tau) 
          \text{ a.e. }\tau\in [t,r]   \big)   \bigg\}
    \end{split}
\end{align}
Then, given a horizon $T$, the definition of the RT is as follows \cite{a008}.
\begin{definition}
	\textbf{Reachable tube (RT):}
	\begin{align}
		R_K(T)=\left\{ s\in\mathbb{R}^n|\exists \tau\in [0,T], \exists a(.)\in \mathscr{A}_0, \right. \nonumber \\ 
		\forall \beta[.] \in \mathfrak{B}_0, 
		\left. \phi_{s,0}^{a,\beta[a]}(\tau) \in K  \right\}
	\end{align}
\end{definition}

In the classical level set method, the viscosity solution of the following HJ PDE of a time-dependent value 
function $V(.,.):\mathbb{R}^n\times \mathbb{R} \to \mathbb{R}$ 
is numerically solved:
\begin{align}
	\left\{ \begin{array}{l}
		\displaystyle{ \frac{\partial V}{\partial t}(s,t)+ 
		\min \left[0, \min_{a\in \mathcal{A}} \max_{b\in\mathcal{B}} \frac{\partial V}{ \partial s}(s,t)f(s,a,b)   \right]=0}\\
		\text{s.t.} V(s,T)=l(s)
	\end{array}\right.
\end{align}
where $l(.)$ is bounded and Lipschitz continuous, and satisfies $K=\left\{s\in\mathbb{R}^n|l(s) \leq 0 \right\}$.
The viscosity solution is approximated using a Cartesian grid of the state space.
The RT can then be characterized by a zero sublevel set of $V(.,0)$:
\begin{align}
	R_K(T)=\left\{s\in\mathbb{R}^n|V(s,0) \leq 0 \right\}
\end{align}
Denote the number of grid points in the $i$th dimension of the Cartesian grid as $N_i$. The storage space consumed to store $R_K(T)$ 
is proportional to $ \prod_{i=1}^n N_i $.

It should be noted that, for $T_1,...,T_M\in [0,\infty)$, 
the expressions of the RTs with these time horizons are as follows:
\begin{align}
	\begin{split}
		&R_K(T_1)=\left\{s\in\mathbb{R}^n|V(s,T-T_1) \leq 0\right\}\\
		&...\\
		&R_K(T_M)=\left\{s\in\mathbb{R}^n|V(s,T-T_M) \leq 0\right\}
	\end{split}
\end{align}

Since the value functions $V(.,T-T_1),...,V(.,T-T_M)$ are each different,
the storage space consumption required to save these RTs is proportional to $ M\prod_{i=1}^n N_i $.

\section{Improved level set method}

% In the rest of this section, we derive a representation of the BRT from another form of the differential 
% game problem and generalize it to the CBRT.

Consider the case where Player I chooses $a(.)$ to enable the system state to reach the target set in as short a time as possible, 
and 
Player II chooses a nonanticipative strategy $\beta[a](.)$
to avoid or delay the entry of the system state into the target set as much as possible. 

If a trajectory can enter the target set within time horizon $T$ in this case, 
then the initial state of this trajectory must belong to the RT. 
Thus, a value function $W(.):\mathbb{R}^n \to \mathbb{R}$ can be constructed as:
\begin{align}
	\label{minvalue}
	W(s)=
		\left\{\begin{array}{rl}
			\displaystyle{\sup_{\beta[.]}}&\displaystyle{ \inf_{a(.) } } \ t_f\\
			\text{s.t.}& \dot{s}(t)=f(s(t),a(t),\beta[a](t))\ \forall t\in [0,t_f]\\
			 \ &s(0)=s\\
			 \ &a(t)\in\mathcal{A} \ \forall t\in [0,t_f]\\
			 \ &\beta[a](t)\in\mathcal{B} \ \forall t\in [0,t_f]\\
			 \ &s(t_f)\in K
		\end{array}\right.
\end{align}
where $t_f$ is the final time and $t_f$ is free.

\begin{remark}\label{rmk0}
It is possible that for some initial states $s$ and Player II's strategies, 
the trajectory initialized from $s$ will never reach the target set no matter what Player I does.
In this case, $W(s)$ is obviously infinity.
\end{remark}

Then, the RT can be characterized by the $T$ sublevel set of $W(.)$, i.e.
\begin{align}
	R_K(T)=\left\{s\in\mathbb{R}^n|W(s) \leq T \right\}
\end{align}

The differential game in Eq. (\ref{minvalue}) 
is a differential game with a terminal state constraint.
To convert it into a differential game without a terminal state constraint 
to facilitate the construction of the value function, 
we need to define a modified dynamic system, i.e.:
\begin{align}
	\label{sys2}
	\dot{s}= \hat{f}(s,a,b)=\begin{cases}
		f(s,a,b), &s\notin K\\
		\mathbf{0}, &s\in K
	\end{cases}
\end{align}
and a modified running cost function, i.e.:
\begin{align}
	\label{runningcost1}
	\mathbb{I}(s) =\begin{cases}
		1, &s\notin K\\
		0, &s\in K
	\end{cases}
\end{align}
Given the initial time $t$ and state $s$, and $a(.)\in \mathscr{A}_t$, and $b(.)\in\mathscr{B}_t$, the evolution of the 
modified system (\ref{sys2}) in the
time interval $[t,\infty)$ can also be expressed as a continuous trajectory $\hat{\phi}_{s,t}^{a,b}(.):[t,\infty) \to \mathbb{R}^n$.
% and the modified performance index is
% \begin{align}
% 	\label{perfidx2}
% 	\hat{\mathcal{J}}_{t_0}^{t_1}(s,a,b) = \int_{t_0}^{t_1} \hat{c}\left( \hat{\phi}_{s,t_0}^{a,b}(t),a(t)    \right) dt
% \end{align}

% \begin{remark}\label{rmk2}
% 	As long as the trajectory $\hat{\phi}_{s,t}^{a,b}(.)$ evolves outside the target set $K$,
% 	it is the same as trajectory $\phi_{s,t}^{a,b}(.)$, 
% 	and the modified running cost function $\hat{c}(s,a)$ is equal to ${c}(s,a)$.
% 	When the trajectory $\hat{\phi}_{s,t}^{a,b}(.)$ 
% 	touches the border of $K$, then it stays on the border under the dynamics (\ref{sys2}), and 
% 	the modified running cost function $\hat{c}(s,a)$ is equal to zero.
% \end{remark}

Then, given a $\bar{T}\in [0,\infty)$, 
a modified value function can be defined based on an differential game problem without a terminal state constraint:
\begin{align}
	\label{minvalue1md}
		W_{\bar{T}}(s,\tau)=
		\left\{\begin{array}{rl}
			\displaystyle{\sup_{\beta[.]}}& \displaystyle{\inf_{a(.)} }  \displaystyle{\int_\tau^{\bar{T}} \mathbb{I}(s(t)) dt} \\
			\text{s.t.}& \dot{s}(t)=\hat{f}(s(t),a(t),\beta[a](t))\ \forall t\in [\tau,\bar{T}]\\
			 \ &s(\tau)=s\\
			 \ &a(t)\in\mathcal{A} \ \forall t\in [\tau,\bar{T}]\\
			 \ &\beta[a](t)\in\mathcal{B} \ \forall t\in [\tau,\bar{T}]
		\end{array}\right.
\end{align}

% With Eq. (\ref{perfidx2}) introduced, the above equation can be simplified to the following form:
% \begin{align}
% 	\label{minvalue2mdsimp}
% 	W_{\bar{T}}^c(s,\tau)=\displaystyle{\sup_{\beta[.]\in\mathfrak{B}}} \displaystyle{\inf_{a(.)\in \mathscr{A}} }  
% 	\displaystyle{\hat{\mathcal{J}}_{\tau}^{\bar{T}}(s,a,\beta[a])}
% \end{align}

Then, we can deduce the following result:

\begin{thm}\label{thm1}
	 $\forall\ \tilde{T} < \bar{T} $ and $s\in \{ s_0\in \mathbb{R}^n| {W}_{\bar{T}}(s_0,0)\leq \tilde{T} \}$, 
	${W}_{\bar{T}}(s,0)=W(s)$.
\end{thm}

% \textbf{Proof:}
\textbf{\textit{Proof:}}
	This theorem is just a special case of a theorem to be given and proved in the next section.

Theorem \ref{thm1} indicates that, 
for $T_1,...,T_M < \gamma \bar{T}$, 
the RTs $R_K(T_1),...,R_K(T_M)$
can be represented as a family of sublevel sets of ${W}_{\bar{T}}(.,0)$ and all RTs can be saved by 
saving ${W}_{\bar{T}}(.,0)$:
\begin{align}
    \begin{split}
        &R_K(T_1)=\left\{s\in\mathbb{R}^n| {W}_{\bar{T}}(s,0) \leq T_1 \right\}\\
        &...\\
        &R_K(T_M)=\left\{s\in\mathbb{R}^n| {W}_{\bar{T}}(s,0) \leq T_M \right\}
    \end{split}
\end{align}

\begin{thm}\label{thm2}
	The value function $W_{\bar{T}}(.,.)$ is the viscosity solution of the following HJ PDE:
	\begin{align}
		\begin{cases}
		\displaystyle{	\frac{\partial W_{\bar{T}}}{\partial t}(s,t)+\min_{a\in\mathcal{A}} \max_{b\in\mathcal{B}} 
		\left\{ \mathbb{I}(s) + \frac{\partial W_{\bar{T}}}{ \partial s }(s,t) \hat{f}(s,a,b)  \right\}=0 }\\
		W_{\bar{T}}(s,\bar{T})=0
		\end{cases}
	\end{align}
\end{thm}

\textbf{\textit{Proof:}}
This theorem is just a special case of another theorem to be given and proved in the next section.

	\section{Generalization of reachability problem}

	\subsection{Definition of a cost-limited Reachable tube}
	As described earlier, a general performance index consists of a Lagrangian. 
	We denote the running cost as $c(.,.,.):\mathbb{R}^n \times \mathcal{A}\times \mathcal{B}\to \mathbb{R}$, and assume that:
	\begin{assumption}\label{ass1}
		$ \displaystyle{\min_{s\in\mathbb{R}^n,a\in\mathcal{A},b\in \mathcal{B}}} c(s,a,b) = \gamma$ holds, where $\gamma$ is a positive real
		number.  
	\end{assumption}
	The performance index of the evolution
	initialized from $s$ at time $t_0$ under control inputs $a(.)$ and $b(.)$ in time interval $[t_0,t_1]$
	is denoted as:
	\begin{align}
		\label{perfidx1}
		\mathcal{J}_{t_0}^{t_1}(s,a,b) = \int_{t_0}^{t_1} c\left( \phi_{s,t_0}^{a,b}(t),a(t),b(t)    \right) dt
	\end{align}
	Then, given a target set $K$ and an admissible cost $J$, CRT can be defined.
	\begin{definition}
		\textbf{Cost-limited Reachable tube (CRT):} 
		\begin{align}
			R^c_K(J)=\left\{ s\in\mathbb{R}^n|\exists \tau\in [0,\infty), 
			\exists a(.) \in \mathscr{A}_0, \right. \nonumber \\ 
			\left. \forall \beta[.] \in \mathfrak{B}_0, 
			  \phi_{s,0}^{a,\beta[a]}(\tau) \in K \land \mathcal{J}_0^\tau(s,a,\beta[a]) \leq J \right\}
		\end{align}
	\end{definition}
	where "$\land$" is the logical operator "AND".
	Informally, under Assumption \ref{ass1}, the performance index $\mathcal{J}_0^t(s_0,a,b)$ increases with an increase in $t$. 
	Furthermore, $R^c_K(J)$ is a set of states that can be transferred into the target set $K$ under any Player II's nonanticipative strategy 
	$\beta[.]$ before the performance index increases to the given admissible cost $J$. 
	
	Consider the case where Player I aims to transfer the system state to the target set with the least possible cost, and Player II plays a nonanticipative strategy to avoid the entry of the system state into the target set or to maximize the cost during state transition. 
	If a trajectory can reach the target set before the performance index increases to the given admissible cost $J$,
	then the initial state of this trajectory must belong to $R^c_K(J)$.
	Similar to Eq. (\ref{minvalue}), a value function can be constructed as follows:
	\begin{align}
		\label{minvalue2}
		W^c(s)=
			\left\{\begin{array}{rl}
				\displaystyle{\sup_{\beta[.]}}& \displaystyle{\inf_{a(.)} } \  \displaystyle{\int_0^{t_f} c(s(t),a(t),\beta[a](t)) dt} \\
				\text{s.t.}& \dot{s}(t)=f(s(t),a(t),\beta[a](t))\ \forall t\in [0,t_f]\\
				 \ &s(0)=s\\
				 \ &a(t)\in\mathcal{A} \ \forall t\in [0,t_f]\\
				 \ &\beta[a](t)\in\mathcal{B} \ \forall t\in [0,t_f]\\
				 \ &s(t_f)\in K
			\end{array}\right.
	\end{align}
	In this case, the value of $W^c(s_0)$ may also be infinity, as stated in Remark \ref{rmk0}. 
	The CRT can be characterized by the $J-$sublevel set of $W^c(.)$:
	\begin{align}
		R^c_K(J)=\left\{s\in\mathbb{R}^n|W^c(s) \leq J \right\}
	\end{align}
	
	\begin{remark}\label{rmk1}
		If $c(s,a,b)\equiv 1$; then, $\displaystyle{\int_0^{t_f} c(s(t),a(t),b(t)) dt}=t_f$. The general performance index
		in Eq. (\ref{minvalue2}) degenerates into the time
		consumption in Eq. (\ref{minvalue}). Similarly, CRT $R^c_K(J)$ is degenerated into RT
		$R_K(J)$.
	\end{remark}

	\subsection{CRT Computation}
	Similar to Eq. (\ref{runningcost1}), the running cost function $c(.,.,.)$ can be modified as follows:
	\begin{align}
		\label{runningcost2}
		\hat{c}(s,a,b) =\begin{cases}
			c(s,a,b), &s\notin K\\
			0, &s\in K
		\end{cases}
	\end{align}

	\begin{remark}\label{rmk2}
		As long as the trajectory $\hat{\phi}_{s,t}^{a,b}(.)$ evolves outside the target set $K$,
		it is the same as trajectory $\phi_{s,t}^{a,b}(.)$, 
		and the modified running cost function $\hat{c}(s,a,b)$ is equal to ${c}(s,a,b)$.
		When the trajectory $\hat{\phi}_{s,t}^{a,b}(.)$ 
		touches the border of $K$, then it stays on the border under the dynamics (\ref{sys2}), and 
		the modified running cost function $\hat{c}(s,a,b)$ is equal to zero.
	\end{remark}
	
	Then, given a $\bar{T}\in [0,\infty)$, 
	a modified value function can be defined based on an differential game problem without a terminal state constraint:
	\begin{align}
		\label{minvalue2md}
			W_{\bar{T}}^c(s,\tau)=
			\left\{\begin{array}{rl}
				\displaystyle{\sup_{\beta[.]}}& \displaystyle{\inf_{a(.)} }  \displaystyle{\int_\tau^{\bar{T}} \hat{c}(s(t),a(t),\beta[a](t)) dt} \\
				\text{s.t.}& \dot{s}(t)=\hat{f}(s(t),a(t),\beta[a](t))\ \forall t\in [\tau,\bar{T}]\\
				 \ &s(\tau)=s\\
				 \ &a(t)\in\mathcal{A} \ \forall t\in [\tau,\bar{T}]\\
				 \ &\beta[a](t)\in\mathcal{B} \ \forall t\in [\tau,\bar{T}]
			\end{array}\right.
	\end{align}
	
	% With Eq. (\ref{perfidx2}) introduced, the above equation can be simplified to the following form:
	% \begin{align}
	% 	\label{minvalue2mdsimp}
	% 	W_{\bar{T}}^c(s,\tau)=\displaystyle{\sup_{\beta[.]\in\mathfrak{B}}} \displaystyle{\inf_{a(.)\in \mathscr{A}} }  
	% 	\displaystyle{\hat{\mathcal{J}}_{\tau}^{\bar{T}}(s,a,\beta[a])}
	% \end{align}
	
	Then, we may deduce the following result:
	
	\begin{thm}\label{thm3}
		For any $\tilde{J} < \gamma \bar{T} $ and any $s\in \{ s_0\in \mathbb{R}^n| {W}_{\bar{T}}^c(s_0,0)\leq \tilde{J} \}$, 
		${W}_{\bar{T}}^c(s,0)=W^c(s)$ holds.
	\end{thm}
	
	% \textbf{Proof:}
	\textbf{\textit{Proof:}}
	According to \textbf{Remark \ref{rmk2}},
	if the trajectory $\hat{\phi}_{s,0}^{a,b}(.)$  cannot reach the target state $K$ within time $\bar{T}$, 
	then its performance index satisfies
	
	\begin{align}
		{W}_{\bar{T}}^c(s,0)\geq \int_0^{\bar{T}} \min_{s\in\mathbb{R}^n,a\in\mathcal{A},b\in\mathcal{B}} c(s,a,b) dt=\gamma \bar{T}
	\end{align}
	
	Therefore, ${W}_{\bar{T}}^c(s,0) \leq \tilde{J} < \gamma \bar{T}$ implies that
	there exists a $a(.)\in \mathscr{A}_0$ and a $\tau\in [0,\bar{T})$ 
	for any $\beta[.]\in \mathfrak{B}_0$, such that $\hat{\phi}_{s,0}^{a,\beta[a]}(\tau)\in K$. 
	In this case, the following equations hold:
	
	\begin{align}
		\begin{split}
		&{W}_{\bar{T}}^c(s,0)=\sup_{\beta[.]\in\mathfrak{B}_0} \inf_{a(.)\in \mathscr{A}_0}\int_{0}^{\bar{T}} 
									\hat{c}\left( \hat{\phi}_{s,0}^{a,\beta[a]}(t),a(t),\beta[a](t)  \right) dt\\
			&=\sup_{\beta[.]\in\mathfrak{B}_0} \inf_{a(.)\in \mathscr{A}_0}\int_{0}^{\tau} 
			\hat{c}\left( \hat{\phi}_{s,0}^{a,\beta[a]}(t),a(t),\beta[a](t)   \right) dt + \int_\tau^{\bar{T}} 0 dt\\
			&=\left\{ \begin{array}{rl}
			\displaystyle{\sup_{\beta[.]\in\mathfrak{B}_0}} &\displaystyle{ \inf_{a(.)\in \mathscr{A}_0}}\int_{0}^{\tau} 
			{c}\left( {\phi}_{s,0}^{a,\beta[a]}(t),a(t),\beta[a](t)   \right) dt\\
			\text{s.t.}& \hat{\phi}_{s,0}^{a,\beta[a]}(\tau)\in K
			\end{array}\right. \\
			&=W^c(s)
		\end{split} 
	\end{align}  \rightline{$\square$}

	\begin{remark}
		Set the running cost function to $c(s,a,b)\equiv 1$, in which case $\hat{c}(s,a,b)=\mathbb{I}(s) $ and $\gamma=1$. By replacing $\tilde{J}$ with $\tilde{T}$, Theorem \ref{thm3} is transformed into Theorem \ref{thm1}.
	\end{remark}

	Theorem \ref{thm3} indicates that, 
	for $J_1,...,J_M < \gamma \bar{T}$, 
	the CRTs $R^c_K(J_1),...,R^c_K(J_M)$
	can be represented as a family of sublevel sets of ${W}_{\bar{T}}^c(.,0)$ and all CRTs can be saved by 
	saving ${W}_{\bar{T}}^c(.,0)$:
	\begin{align}
		\begin{split}
			&R^c_K(J_1)=\left\{s\in\mathbb{R}^n| {W}_{\bar{T}}^c(s,0) \leq J_1 \right\}\\
			&...\\
			&R^c_K(J_M)=\left\{s\in\mathbb{R}^n| {W}_{\bar{T}}^c(s,0) \leq J_M \right\}
		\end{split}
	\end{align}
	
	\begin{lemma}
		Given $0 \leq t<t+\Delta t\leq \bar{T} $ and $s\in \mathbb{R}^n$, 
		\begin{align}
			\label{th1eq0}
				W_{\bar{T}}^c(s,t)=\sup_{\beta[.]} \inf_{a(.)} \left\{\int_t^{t+\Delta t} 
				\hat{c}\left(\hat{\phi}_{s,t}^{a,\beta[a]}(\tau), a(\tau),\beta[a](\tau) \right) d\tau \right. \nonumber \\
				\left. +W_{\bar{T}}^c\left(\hat{\phi}_{s,t}^{a,\beta[a]}(t+\Delta t),t+\Delta t\right)\right\}
		\end{align}
	\end{lemma}
	
	\textbf{\textit{Proof:}}
		Let 
		\begin{align}
			\overline{W}_{\bar{T}}^c(s,t)=\sup_{\beta[.]} \inf_{a(.)} \left\{\int_t^{t+\Delta t} 
				\hat{c}\left(\hat{\phi}_{s,t}^{a,\beta[a]}(\tau), a(\tau),\beta[a](\tau) \right) d\tau \right. \nonumber \\
				\left.
				+W_{\bar{T}}^c\left(\hat{\phi}_{s,t}^{a,\beta[a]}(t+\Delta t),t+\Delta t\right)\right\}
		\end{align}
		and fix $\epsilon>0$. Then there exists 
		$\bar{\beta}[.]\in \mathfrak{B}_t $ such that 
		\begin{align}
			\overline{W}_{\bar{T}}^c(s,t)\leq 
			 \inf_{a(.)} \left\{\int_t^{t+\Delta t} 
				\hat{c}\left(\hat{\phi}_{s,t}^{a,\bar{\beta}[a]}(\tau), a(\tau),\bar{\beta}[a](\tau) \right) d\tau \right. \nonumber \\
				\left.
				+W_{\bar{T}}^c\left(\hat{\phi}_{s,t}^{a,\beta[a]}(t+\Delta t),t+\Delta t\right)\right\}+\epsilon
		\end{align}
		Also, for each $s'\in \mathbb{R}^n$
		\begin{align}
			\label{th1eq1}
			&{W}_{\bar{T}}^c(s',t+\Delta t)= \nonumber \\
			&\sup_{\beta[.]\in\mathfrak{B}_{t+\Delta t}} \inf_{a(.)\in \mathscr{A}_{t+\Delta t}}
			\int_{t+\Delta t}^{\bar{T}} 
									\hat{c}\left( \hat{\phi}_{s',t+\Delta t}^{a,\beta[a]}(\tau),a(\tau),\beta[a](\tau)  \right) d\tau
		\end{align}
		Thus there exists $\bar{\beta}'[.]\in \mathfrak{B}_{t+\Delta t} $ for which
		\begin{align}
			\label{th1eq2}
			&{W}_{\bar{T}}^c(s',t+\Delta t)\leq \nonumber \\
			& \inf_{a(.)\in \mathscr{A}_{t+\Delta t}}
			\int_{t+\Delta t}^{\bar{T}} 
									\hat{c}\left( \hat{\phi}_{s',t+\Delta t}^{a,\bar{\beta}'[a]}(\tau),a(\tau),\bar{\beta}'[a](\tau)  \right) d\tau
									+\epsilon
		\end{align}
		Define $\beta[.]\in \mathfrak{B}_t$ in this way: for each $a(.)\in \mathscr{A}_t$ set 
		\begin{align}
			\beta[a](\tau)=\begin{cases}
				\bar{\beta}[a](\tau), &\tau \in [t,t+\Delta t]\\
				\bar{\beta}'[a](\tau), &\tau \in (t+\Delta t,\bar{T}]
			\end{cases}
		\end{align}
		Consequently, for any $a(.)\in \mathscr{A}_t$, Eq. (\ref{th1eq1}) and Eq. (\ref{th1eq2}) imply
		\begin{align}
			\overline{W}_{\bar{T}}^c(s,t)\leq  \int_t^{\bar{T}} 
				\hat{c}\left(\hat{\phi}_{s,t}^{a,\beta[a]}(\tau), a(\tau),\beta[a](\tau) \right) d\tau
				+2\epsilon
		\end{align}
		So that
		\begin{align}
			\inf_{a(.)} \int_t^{\bar{T}}\hat{c}\left(\hat{\phi}_{s,t}^{a,\beta[a]}(\tau), a(\tau),\beta[a](\tau) \right)d\tau
			\geq \overline{W}_{\bar{T}}^c(s,t)-2\epsilon
		\end{align}
		Hence
		\begin{align}
			\label{th1eq2.4}
			{W}_{\bar{T}}^c(s,t) \geq \overline{W}_{\bar{T}}^c(s,t)-2\epsilon
		\end{align}
		On the other hand, there exists $\tilde{\beta}[.]\in \mathfrak{B}_t$ for which 
		\begin{align}
			\label{th1eq2.5}
			{W}_{\bar{T}}^c(s,t)\leq \inf_{a(.)} 
			\int_t^{\bar{T}}\hat{c}\left(\hat{\phi}_{s,t}^{a,\tilde{\beta}[a]}(\tau), a(\tau),\tilde{\beta}[a](\tau) \right)d\tau
			+\epsilon
		\end{align}
		Then 
		\begin{align}
			\overline{W}_{\bar{T}}^c(s,t)\geq \inf_{a(.)}
			\left\{ \int_t^{t+\Delta t}\hat{c}\left(\hat{\phi}_{s,t}^{a,\tilde{\beta}[a]}(\tau), a(\tau),\tilde{\beta}[a](\tau) \right)d\tau 
			\right. \nonumber \\ 
			\left.+ W_{\bar{T}}^c\left(\hat{\phi}_{s,t}^{a,\beta[a]}(t+\Delta t),t+\Delta t\right) \right\}
		\end{align}
		and consequently there exists $\bar{a}(.)\in\mathscr{A}_t$ such that 
		\begin{align}
			\label{th1eq3}
			\overline{W}_{\bar{T}}^c(s,t)\geq 
			\left\{ \int_t^{t+\Delta t}\hat{c}\left(\hat{\phi}_{s,t}^{\bar{a},\tilde{\beta}[a]}(\tau), 
			\bar{a}(\tau),\tilde{\beta}[\bar{a}](\tau) \right)d\tau \right. \nonumber \\
			\left. + W_{\bar{T}}^c\left(\hat{\phi}_{s,t}^{a,\beta[a]}(t+\Delta t),t+\Delta t\right) \right\}-\epsilon
		\end{align}
		Now define $\tilde{a}(.)\in \mathscr{A}_t$ by 
		\begin{align}
			\tilde{a}(\tau)=\begin{cases}
				\bar{a}(\tau), &\tau\in [t,t+\Delta t]\\
				a(\tau),&\tau\in (t+\Delta t,\bar{T}]
			\end{cases}
		\end{align}
		and then define $\tilde{\beta}'\in \mathfrak{B}_{t+\Delta t}$ by
		\begin{align}
			\tilde{\beta}'[a](\tau)=\tilde{\beta}[\tilde{a}](\tau)
		\end{align}
		Hence
		\begin{align}
			&W_{\bar{T}}^c\left(\hat{\phi}_{s,t}^{a,\beta[a]}(t+\Delta t),t+\Delta t\right) 
			\geq \nonumber \\
			&\inf_{a(.)\in \mathscr{A}_{t+\Delta t}} \int_{t+\Delta t}^{\bar{T}}
			\hat{c}\left(\hat{\phi}_{s,t}^{{a},\tilde{\beta}'[a]}(\tau), {a}(\tau),\tilde{\beta}'[{a}](\tau) \right)d\tau
		\end{align}
		and there exists $\tilde{a}'\in \mathscr{A}_{t+\Delta t}$ for which
		\begin{align}
			\label{th1eq4}
			&W_{\bar{T}}^c\left(\hat{\phi}_{s,t}^{\tilde{a}',\beta[\tilde{a}']}(t+\Delta t),t+\Delta t\right) 
			\geq \nonumber \\
			&\int_{t+\Delta t}^{\bar{T}}
			\hat{c}\left(\hat{\phi}_{s,t}^{{\tilde{a}'},\tilde{\beta}'[\tilde{a}']}(\tau), \tilde{a}'(\tau),\tilde{\beta}'[\tilde{a}'](\tau) \right)d\tau
			-\epsilon
		\end{align}
		Set $a(.)\in \mathscr{A}_t$ as 
		\begin{align}
			a(\tau)=\begin{cases}
				\tilde{a}(\tau), &\tau\in [t,t+\Delta t]\\
				\tilde{a}'(\tau), &\tau \in (t+\Delta, \bar{T} ]
			\end{cases}
		\end{align}
		Then Eq. (\ref{th1eq3}) and Eq. (\ref{th1eq4}) yield 
		\begin{align}
			\overline{W}_{\bar{T}}^c(s,t)\geq \int_t^{\bar{T}} 
			\hat{c}\left(\hat{\phi}_{s,t}^{{a},\tilde{\beta}[a]}(\tau), a(\tau),\tilde{\beta}[a](\tau) \right)d\tau-2\epsilon
		\end{align}
		Therefore, Eq. (\ref{th1eq2.5}) implies
		\begin{align}
			\label{th1eq5}
			{W}_{\bar{T}}^c(s,t) \leq \overline{W}_{\bar{T}}^c(s,t)+3\epsilon
		\end{align}
		Eqs. (\ref{th1eq5}) (\ref{th1eq2.4}) complete the proof. \rightline{$\square$}

	\begin{thm}\label{thm4}
		The value function ${W}_{\bar{T}}^c(.,.)$ is the viscosity solution of the following HJ PDE:
		\begin{align}
			\label{hjpde}
			\begin{cases}
				\displaystyle{	\frac{\partial W_{\bar{T}}^c}{\partial t}(s,t)+\min_{a\in\mathcal{A}} \max_{b\in\mathcal{B}} 
				\left\{ \hat{c}(s,a,b) + \frac{\partial W_{\bar{T}}^c}{ \partial s }(s,t) \hat{f}(s,a,b)  \right\}=0 }\\
				W_{\bar{T}}^c(s,\bar{T})=0
			\end{cases}
		\end{align}
	\end{thm}
	
	\textbf{\textit{Proof:}}
		A nonanticipative strategy is equivalent to allowing Player II to choose $b(t)$ based on knowledge of $a(\tau)$ 
		for all $\tau\in [t,\bar{T}]$; 
		in other words, Player I would have to reveal her entire input signal in advance to Player II \cite{a008}.
		Therefore, for a small enough $\Delta t$, Eq. (\ref{th1eq0}) can be rewritten as:
		\begin{align}
			\label{th2eq1}
			W_{\bar{T}}^c(s,t)=\min_{a\in\mathcal{A}} \max_{b\in\mathcal{B}} \bigg\{
				\hat{c}\left( s,a,b \right) \Delta t  \nonumber \\
				+W_{\bar{T}}^c\left(s+\hat{f}(s,a,b)\Delta t ,t+\Delta t\right)\bigg\}
		\end{align}
		The Taylor expansion of $W_{\bar{T}}^c(.,.)$ at $(s,t)$ yields
		\begin{align}
			\label{th2eq2}
			&W_{\bar{T}}^c\left(s+\hat{f}(s,a,b)\Delta t ,t+\Delta t\right)= \nonumber \\
			&W_{\bar{T}}^c\left(s ,t\right)+\frac{\partial W_{\bar{T}}^c}{\partial s}(s,t) \hat{f}(s,a,b)\Delta t
			+\frac{\partial W_{\bar{T}}^c}{\partial t}(s,t) \Delta t
		\end{align}
		Substituting Eq. (\ref{th2eq2}) into Eq. (\ref{th2eq1}) yields 
		\begin{align}
			\frac{\partial W_{\bar{T}}^c}{\partial t}(s,t)+\min_{a\in\mathcal{A}} \max_{b\in\mathcal{B}} 
				\left\{ \hat{c}(s,a,b) + \frac{\partial W_{\bar{T}}^c}{ \partial s }(s,t) \hat{f}(s,a,b)  \right\}=0
		\end{align}
		On the other hand, 
		\begin{align}
			\label{th2eq3}
			W_{\bar{T}}^c(s,\bar{T})=\int_{\bar{T}}^{\bar{T}} \hat{c}(s(t),a(t),\beta[a](t)) dt=0
		\end{align}
		Eq. (\ref{th2eq2}) and Eq. (\ref{th2eq3}) complete the proof. \rightline{$\square$}

	\begin{remark}
		Set the running cost function to $c(s,a,b)\equiv 1$, in which case $\hat{c}(s,a,b)=\mathbb{I}(s) $ and $W_{\bar{T}}^c(.,.)$ degenerate into $W_{\bar{T}}(.,.)$. Therefore, Theorem \ref{thm2} is a special case of Theorem \ref{thm4}.
	\end{remark}

	Theorem \ref{thm3} and Theorem \ref{thm4} imply that for any $J\leq \gamma \bar{T}$, the $J$-sublevel set of 
	the solution of the HJ PDE in Eq. (\ref{hjpde}) describes $R^c_K(J)$.
	The work \cite{levelsetbook} presents in detail the viscosity solution of the HJ PDE.
	It is first necessary to specify a rectangular region in the state space as the computational domain,
	and then discretize it into a Cartesian grid structure.
	The accuracy of the viscosity solution depends on the number of grids.

	The pseudocode of the proposed method is shown in Algorithm 1.
	
	\begin{algorithm}[h]
		\caption{Method to compute CRTs}
		\label{alg1}
		\begin{algorithmic}[1]
			\State \textbf{Inputs:} Dynamic system (\ref{sys1}), sets of players' achievable control inputs $\mathcal{A}$ and $\mathcal{B}$, 
			running cost function $c(.,.,.)$,
			admissible costs $J_1,...,J_M$, target set $K$, 
			computational domain $\Omega$, 
			number of grids $N_x\times N_y\times ...$; 
			\State $\displaystyle{ \gamma\leftarrow \min_{s\in\mathbb{R}^n,a\in\mathcal{A},b\in\mathcal{B}} c(s,a,b) }$, 
	
			\State $J_{\text{max}}\leftarrow\max\left( J_1,...,J_M \right)$, 
			$\displaystyle{\bar{T}\leftarrow \frac{J_{\text{max}}}{\gamma}+\epsilon} $; 
			\textcolor{gry}{$\backslash\backslash$ $\epsilon$ 
			is a small positive number to ensure $\gamma\bar{T} > J_{\text{max}} $.}
	
			\State Solve the viscosity solution $W_{\bar{T}}^c(.,.) $ of Eq. (\ref{hjpde});
		
			\State $R^c_K(J_1) \leftarrow \left\{s\in\mathbb{R}^n|W_{\bar{T}}^c(s,0)\leq J_1 \right\},...,
			R^c_K(J_M)\leftarrow \left\{s\in\mathbb{R}^n|W_{\bar{T}}^c(s,0)\leq J_M \right\}$;
			\State \textbf{Return} $R^c_K(J_1),..,R^c_K(J_M)$;
		\end{algorithmic}
	\end{algorithm}

	\section{Numerical Examples}

	In this section, two examples are presented to illustrate the validity of the proposed method. 
	The first example shows the RT of a 2D linear system, which can be analytically solved, 
	to analyze the effects of the grid number on computational accuracy.
	The second example, on a pursuit-evasion game, is  
	a practical application of the CRT.

	\subsection{Two-dimensional linear system}
	Consider the following system:
	\begin{align}
		\label{sys4}
		\dot{s}=\left[
			\begin{array}{c}
				\dot{x}\\ \dot{y}
			\end{array}\right]=\left[
				\begin{array}{c}
					b\\-x+a
				\end{array}
			\right]
	\end{align}
	%%%%%%%%%%%%%%%强调二维插值
	where $s=[x,y]^\mathrm{T}\in \mathbb{R}^2$ is the system state, $a\in\mathcal{A}=[-1,1]$ is the control input of Player I, and
	$b\in\mathcal{B}= [-1,1]$ is the control input of Player II. The target set is $K=\left\{[x,y]^\mathrm{T}| |y|\leq 0.5 \right\}$,
	the time horizon is $T=1$. To compute the RT, the running cost function is set to 
	$c(s,a,b)\equiv 1$. 
	The analytical solution for this problem is as follows:
	\begin{align}
		&R^*_K(\bar{T})=K\cup \{[x,y]^\mathrm{T} |0\leq y\leq x+1  \} \nonumber\\
		&\cup \{[x,y]^\mathrm{T} |0\geq y\geq x-1 \}\nonumber \\
		 &\cup \{[x,y]^\mathrm{T} |0\leq y \leq \frac{1}{2}x^2+x+1,-1\leq x\leq 0  \} \nonumber\\
		 &\cup \{[x,y]^\mathrm{T} |0\geq y \geq -\frac{1}{2}x^2+x-1,0\leq x\leq 1  \}
	\end{align}

\begin{figure}[h]
		\centering
		\includegraphics[width=0.375\textwidth]{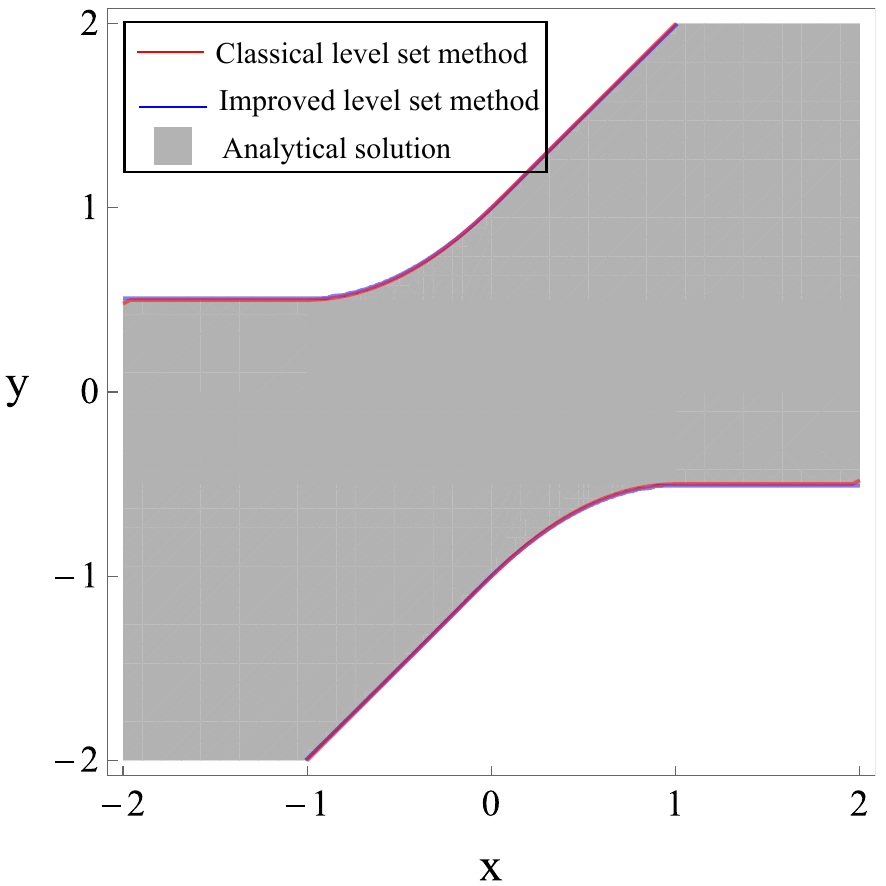}
		\caption{Results of our method and the classical level set method.}
		\label{fig3}
	\end{figure}

	To analyze the convergence of the proposed method, we compute the RT with several sets of parameters
	and compare the results with those of the level set method. 
	Table \ref{tb1} outlines the parameters that are specified for the RT computation,
	and the result for one set of parameters ($N_x,N_y = 251$) are shown in 
	Fig. \ref{fig3}, which also displays the result obtained with the level set method for the same computational domain and grid number.
	It can be seen that the results of the proposed method and the level 
	set method almost coincide, and both are very close to the analytical solution,
	which indicates that the proposed method is highly accurate.
	
	\begin{table}[h]
		\centering
		\caption{Solver settings for the two-dimensional system example}
		\label{tb1}
		\begin{tabular}{lc}
		\toprule[1pt]  
		\textbf{Parameter} 			  			& \textbf{Setting}  \\ \hline
		Computational domain $\Omega$         	&      $[-2,2]\times [-2,2]$      \\   
		Grid points $N_x, N_y$   				&  $51$, $101$, $151$, $201$, $251$     \\  
		$\bar{T}$               				&      $1.2$      \\     \bottomrule[1pt] 
		\end{tabular}
	\end{table}
	
	\begin{figure}[h]
		\centering
		\includegraphics[width=0.4\textwidth]{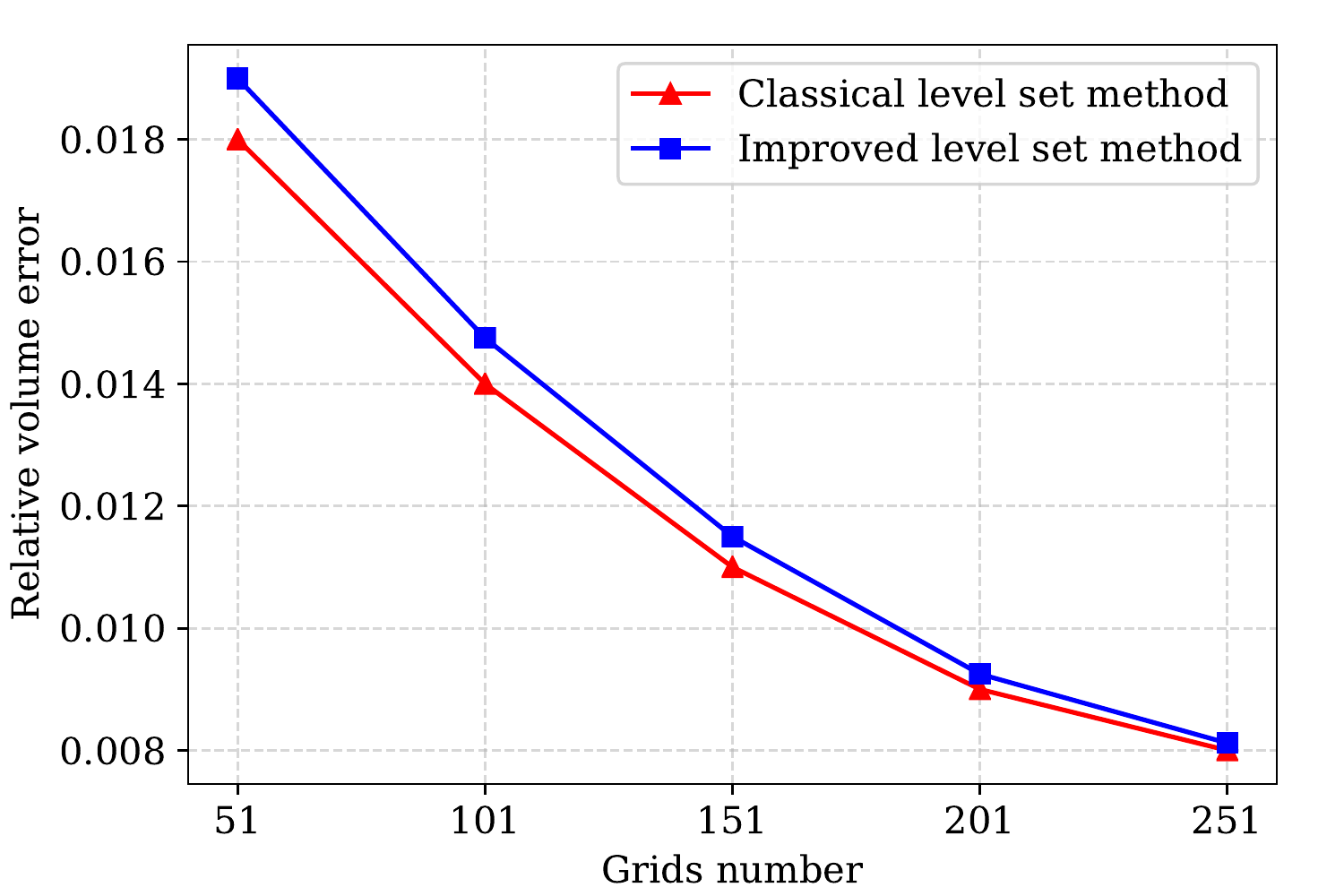}
		\caption{Variations in the relative volume errors with the number of grids.}
		\label{fig4}
	\end{figure}
	
	To quantify the computational errors between the computational results and the analytical solution under different parameters,
	the relative volume errors is quantified using the Jaccard index \cite{jaccard}. 
	In this study, the expression for the relative volume error between set $X$ and set $Y$ is: 
	\begin{align}
		e_{\text{VOL}}(X,Y)=1-\frac{ |X\cap Y |  }{|X\cup Y|}
	\end{align}

	The variations in the relative volume errors with the number of grids are shown in Fig. \ref{fig4}.
	It can be seen that the variation in the computational accuracy of 
	the proposed method with the number of grids is comparable to that of the level set method, and 
	there is no significant effect of the time step size on the computational accuracy.

\begin{figure}[htbp]
		\centering
		\subfigure[The improved level set method.]{\includegraphics[width=0.5\textwidth]{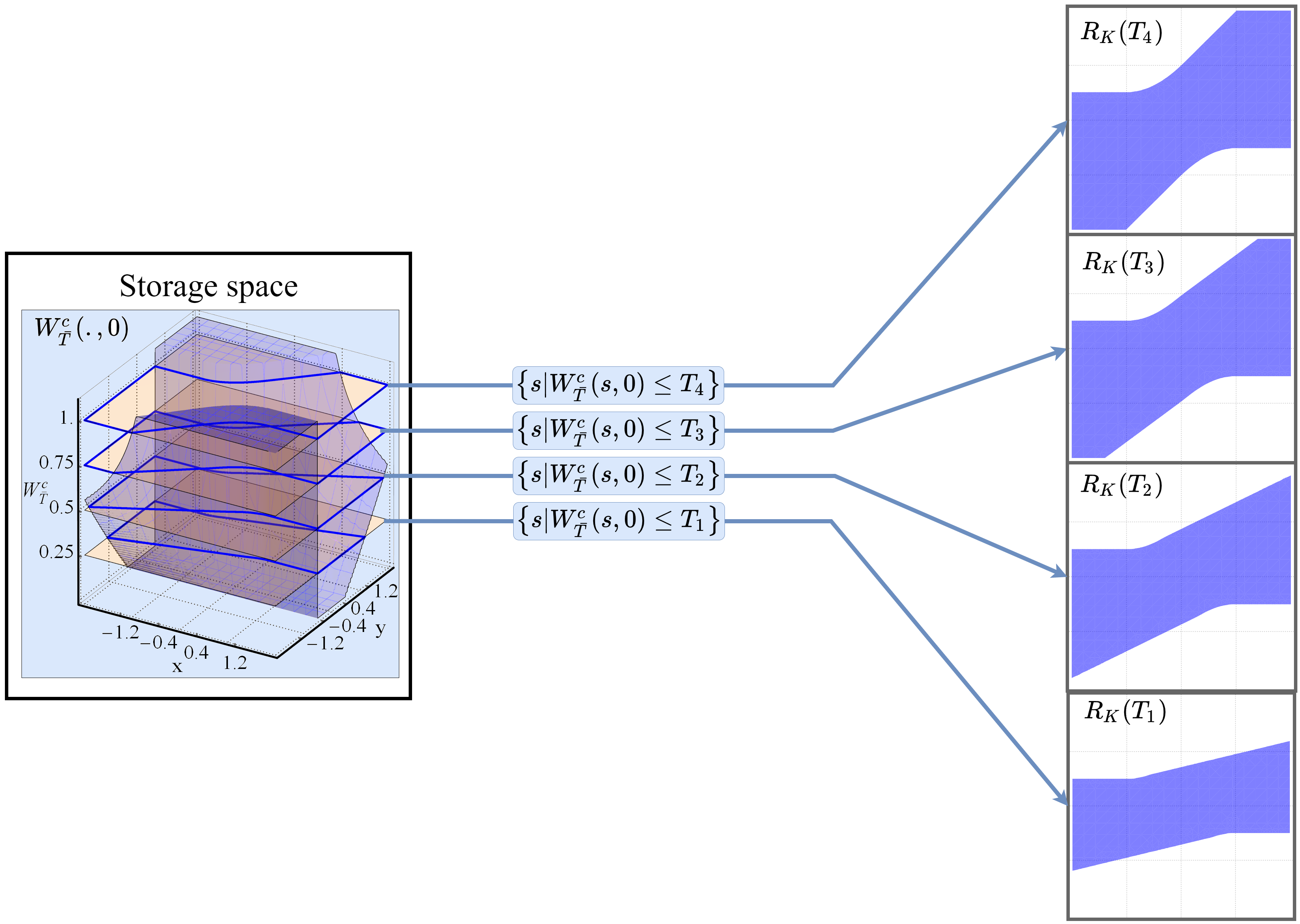}}\\
		\subfigure[The classical level set method.]{\includegraphics[width=0.5\textwidth]{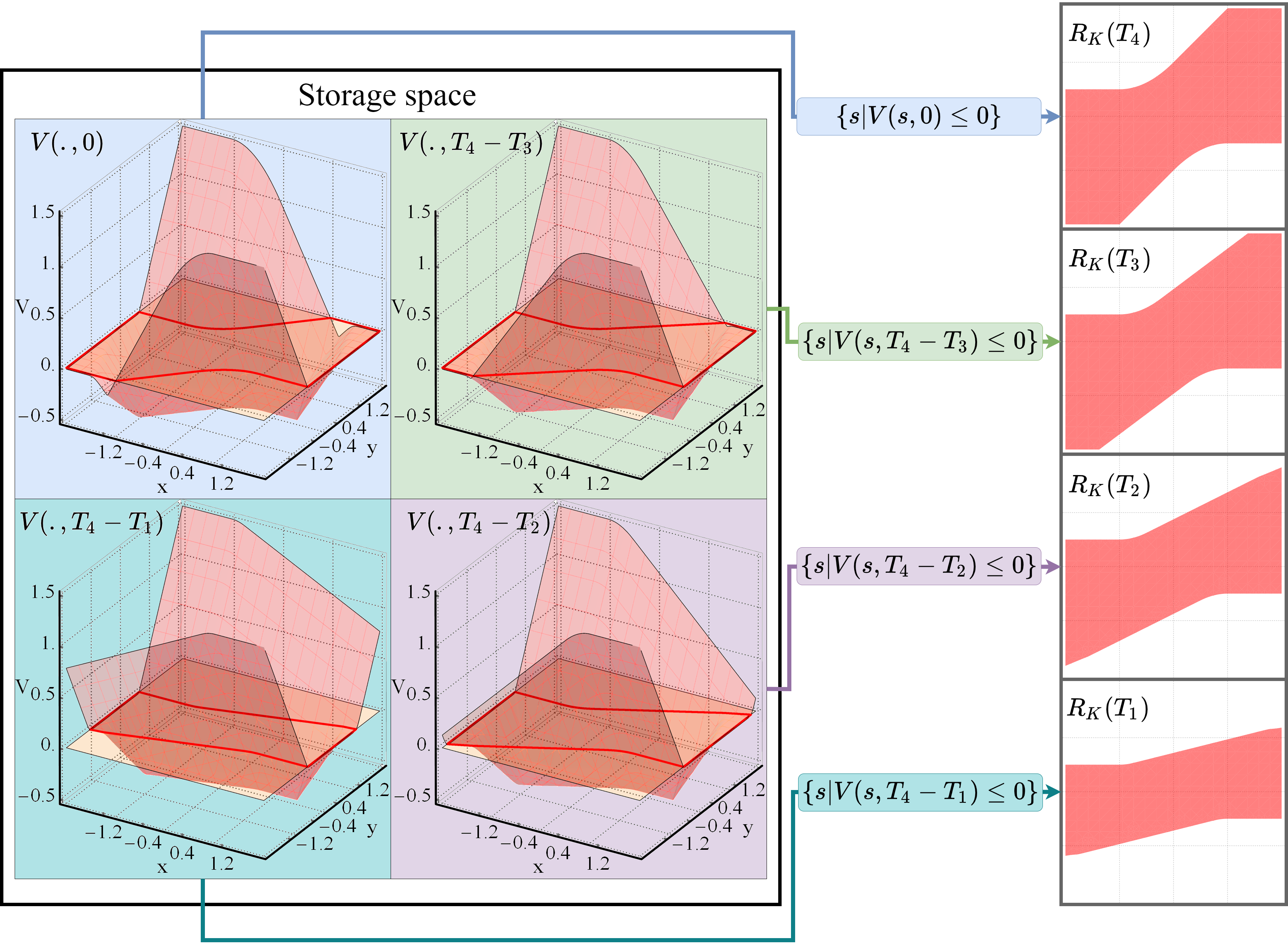}}
		\caption{Visualization of storage forms.}
		\label{fig5}
	\end{figure}

	Although the proposed method is not superior to the level set method in terms of accuracy, 
	it has a significant advantage in terms of storage space consumption. 
	Taking the four time horizons $T_1=0.25$, $T_2=0.5$, $T_3=0.75$, and $T_4=1$ as an example,
	in the proposed method, only $W_{\bar{T}}^c(.,0)$ needs to be saved to save all four RTs 
	$R_K(T_1)$, $R_K(T_2)$, $R_K(T_3)$, and $R_K(T_4)$, see Fig. \ref{fig5}(a).
	In contrast, in the classical level set method, 
	if the terminal condition of HJ PDE is set to $V(s,T_4)=l(s)$,  
	in order to save all four RTs, one needs to save $V(.,T_4-T_1)$, $V(.,T_4-T_2)$, 
	$V(.,T_4-T_3)$, and $V(.,0)$, see Fig. \ref{fig5}(b). 
	The proposed method consumes only a quarter of the storage space of the level set method for the same number of grids.
	
	\subsection{The pursuit-evasion game}
\begin{figure}[h]
		\centering
		\includegraphics[width=0.4\textwidth]{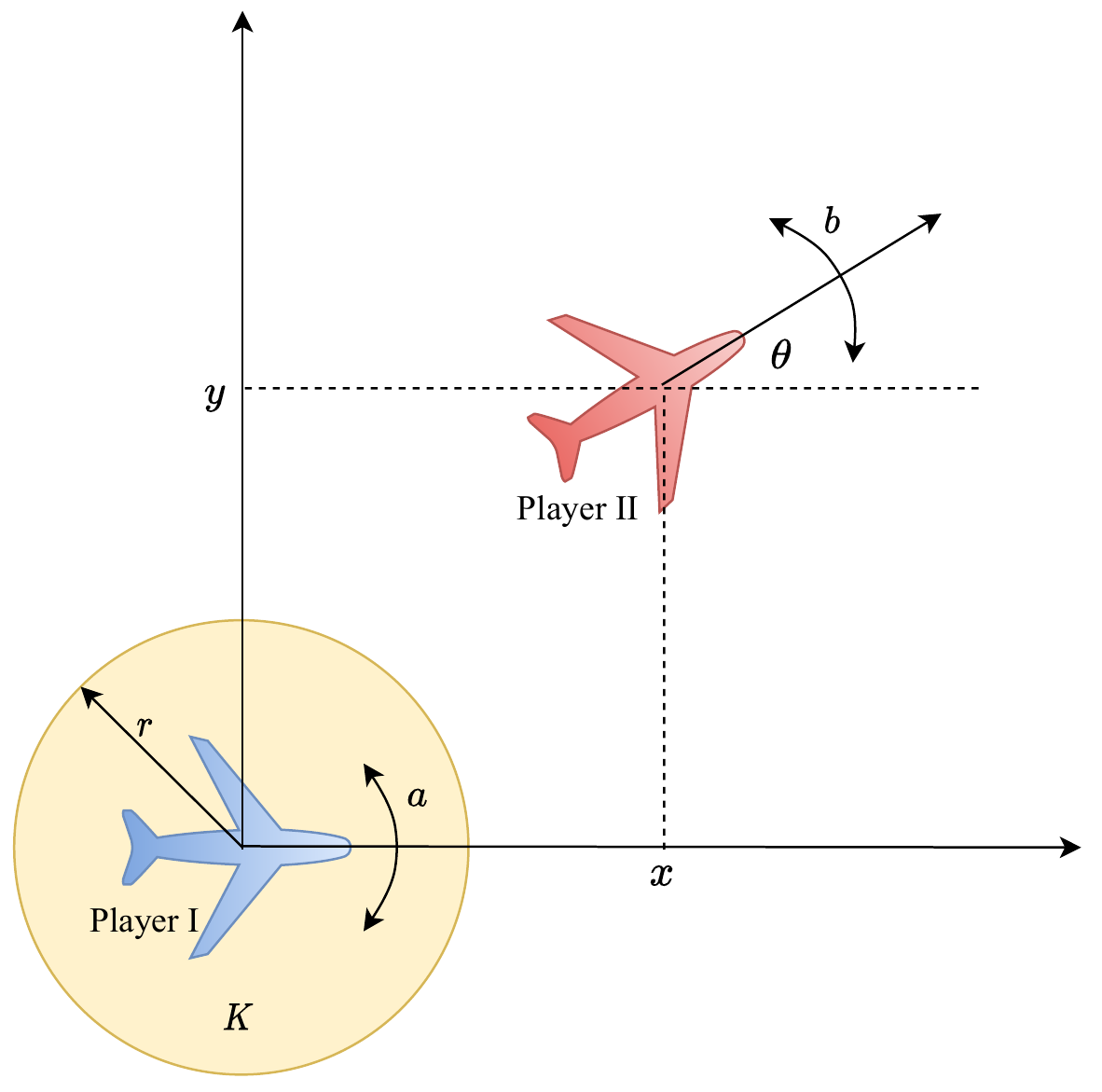}
		\caption{Visualization of the pursuit-evasion game.}
		\label{figpegame}
	\end{figure}

	Two flight vehicles move on a plane without obstacles. 
	Both vehicles are modeled as a simple mass point with fixed linear velocity $v$ and controllable
	angular velocity.
	The pursuer, Player I, tries to capture the evader while the evader, Player II, tries to get away from the pursuer.
	When the distance between two vehicles is less than $r$, we say that a collision has occurred. The task
	is to determine the set of initial states from
	which the pursuer can cause a collision to occur within the given admissible cost.
	Translating into reachability terms, $s=[x,y,\theta]^\mathrm{T}\in \mathbb{R}^2 \times [0,2\pi]$ is the system state, 
	the target set $K=\{[x,y,\theta]^\mathrm{T}| x^2+y^2\leq r^2\}$.
	The evolution of the system can be described by the following equation:
	\begin{align}\left[
		\begin{array}{c}
			\dot{x}\\ \dot{y}\\ \dot{\theta}
		\end{array}\right]=
		\left[\begin{array}{c}
			-v+v\cos \theta +ay\\
			v\sin \theta-ax\\
			b-a
		\end{array}
		\right]
	\end{align}
	where $a\in \mathcal{A}= [-1,1]$ and $b\in \mathcal{B}= [-1,1]$ are control inputs of 
	Player I and Player II, respectively.
	The variables in the preceding equation are described in Fig. \ref{figpegame}.
	
	The parameters in this example are set as:
	\begin{align}
		v=5,r=5
	\end{align}
	The running cost function is 
	% a weighted sum of the time consumption and 
	% maneuver overload factor, i.e.:
	\begin{align}
		c(s,a,b)=1+\lambda (x^2+y^2+a^2+b^2)
	\end{align}
	and the given admissible costs 
	are $J_1=0.5,J_2=1,J_3=1.5$, and $J_4=2$.
	
	% In the preceding equation, $\lambda_G$ is the weight of the maneuver overload,
	% $F_x$ and $F_z$ are the components of the summation of the aerodynamic force and thrust in the horizontal and vertical directions, 
	% respectively:
	% \begin{align}
	%     \begin{split}
	%     &F_x=P\cos\theta-\frac{1}{2}\rho v^2 S \left[ C_L \sin(\theta-\alpha)- C_D \cos(\theta-\alpha) \right]\\
	%     &F_z=P\sin\theta+\frac{1}{2}\rho v^2 S\left[ C_L \cos(\theta-\alpha)- C_D \sin(\theta-\alpha)\right]
	%     \end{split}
	% \end{align}
	
	We consider two cases: $\lambda=0$ and $\lambda=0.1$. In the former case, the running cost $c(s,a,b)\equiv 1$
	such that the problem degenerates to the computation of RT, which can be solved by the classical level set method. 
	The solver settings of our method are listed in Table \ref{tb3}.
	\begin{table}[h]
		\centering
		\caption{Solver settings for the pursuit-evasion game}
		\label{tb3}
		\begin{tabular}{lc}
		\toprule[1pt] 
		\textbf{Parameter} 			  			& \textbf{Setting}  \\ \hline
		Computational domain $\Omega$         	&       $[-5,20]\times[-10,10]\times[0,2\pi]$     \\   
		Grid points $N_x\times N_y\times N_\theta$   &  	 $101\times 101 \times 101$     \\  
		$\bar{T}$               				&      $2.2$      \\     \bottomrule[1pt] 
		\end{tabular}
	\end{table}
	
\begin{figure}[h]
		\centering
		\subfigure[$R_K^c(J_1)$]{\includegraphics[width=0.24\textwidth]{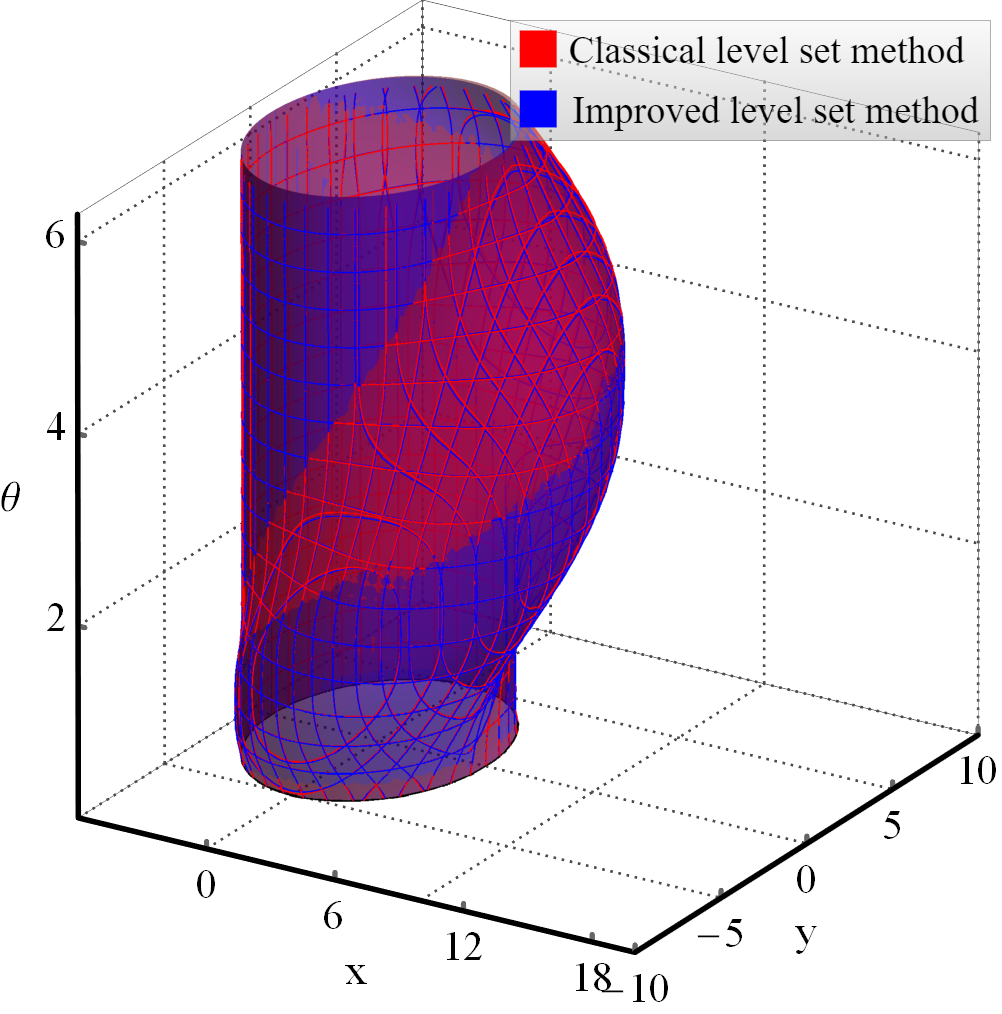}} 
		\subfigure[$R_K^c(J_2)$]{\includegraphics[width=0.24\textwidth]{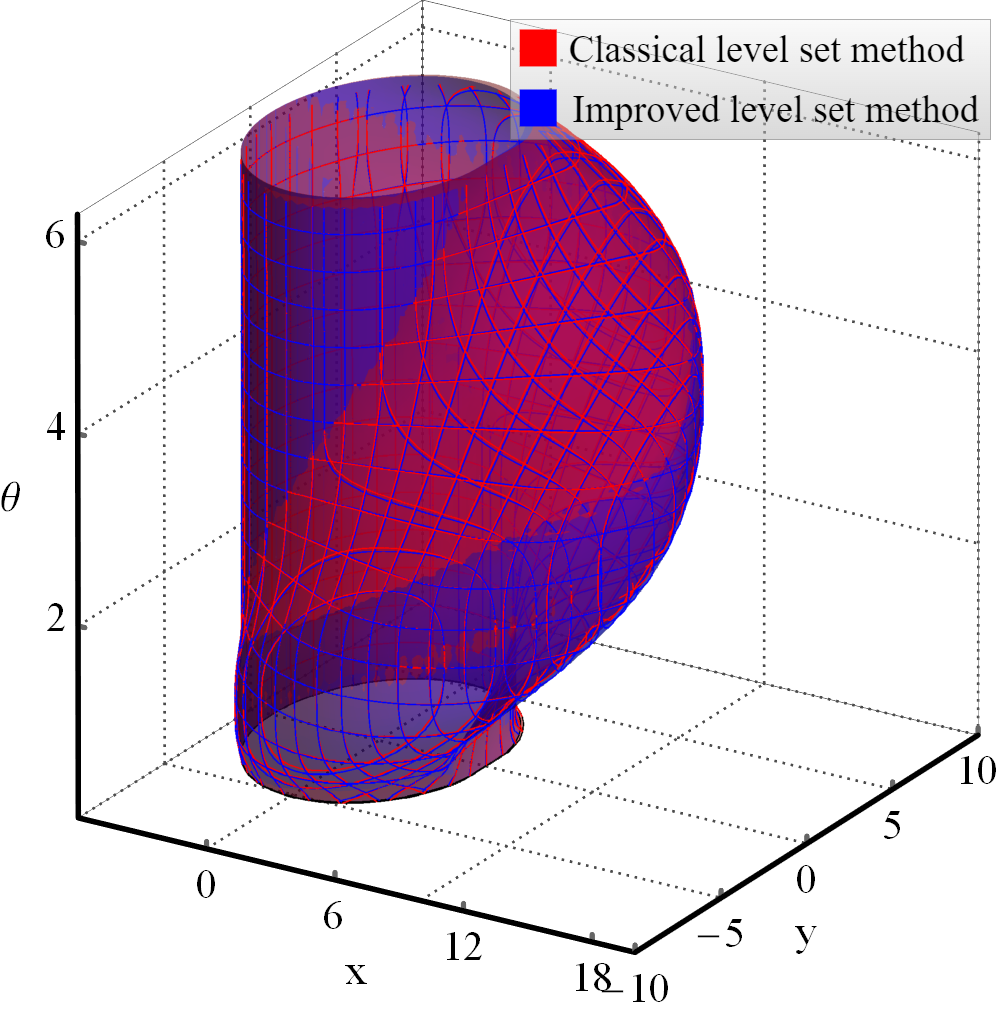}}\\
		\subfigure[$R_K^c(J_3)$]{\includegraphics[width=0.24\textwidth]{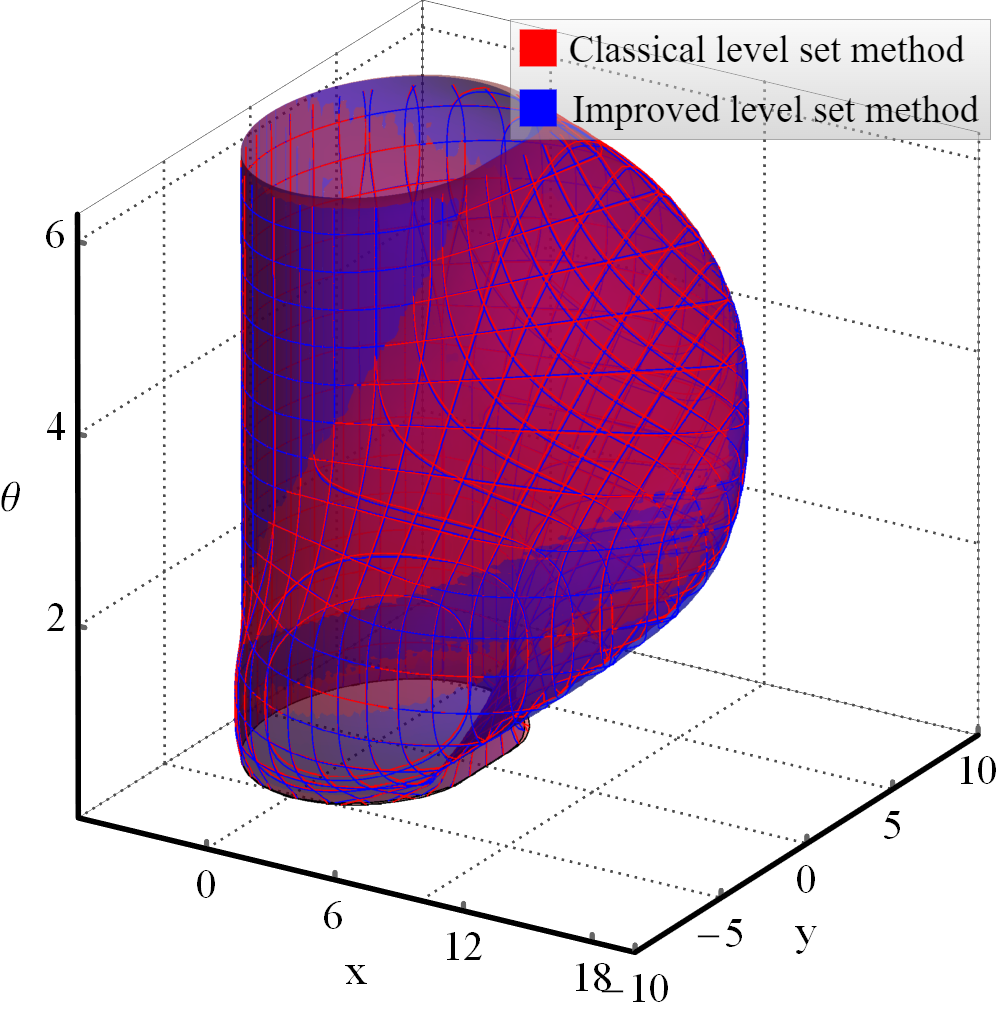}} 
		\subfigure[$R_K^c(J_4)$]{\includegraphics[width=0.24\textwidth]{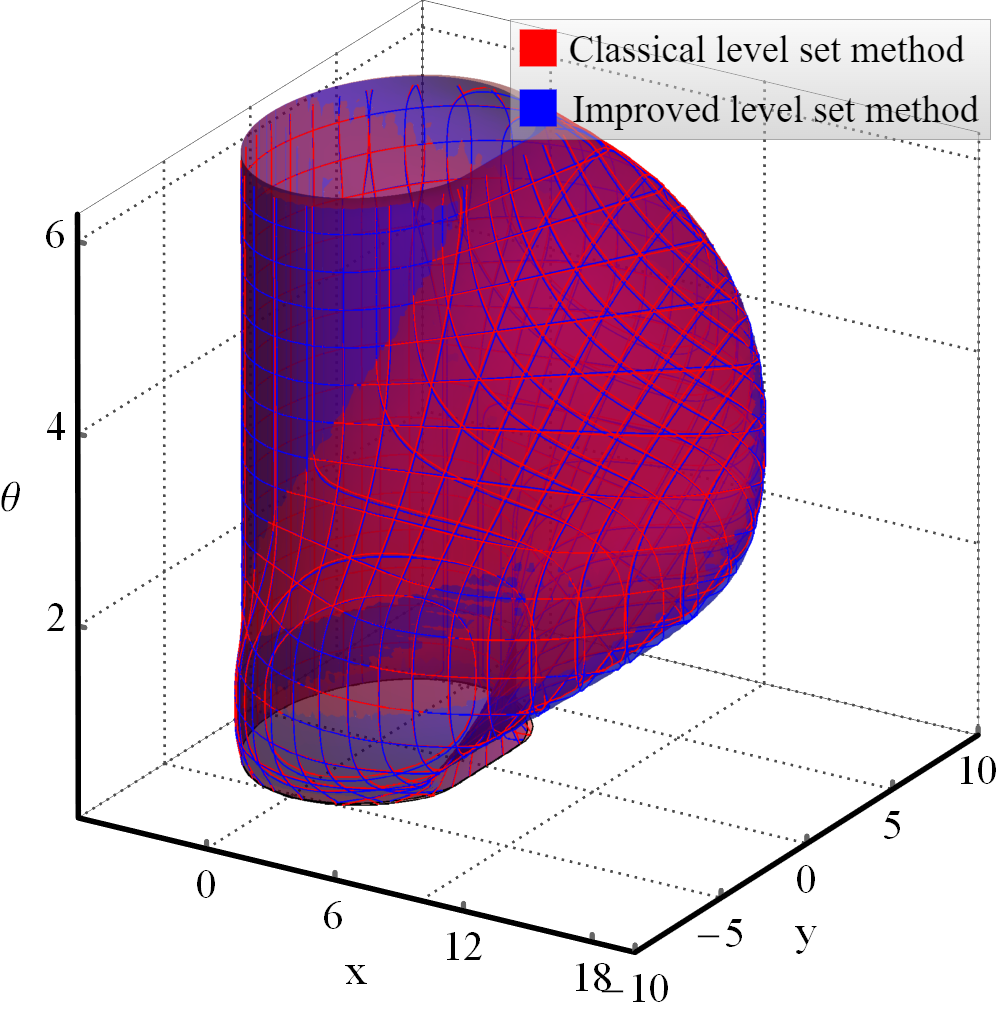}}
		\caption{Computation results of the CRTs with $\lambda=0$.}
		\label{figexp21}
	\end{figure}

	A comparison between the results of the proposed method and the classical level set method is shown in Fig. \ref{figexp21}.
	The setups of the computational domain and the number of grid points in the classical level set method are the same as those in our method.
	It can be seen that, the envelope of the RT computed by the proposed method almost 
	coincides with that computed using the classical level set method.

	\begin{figure}[h]
		\centering
		\subfigure[$R_K^c(J_1)$]{\includegraphics[width=0.24\textwidth]{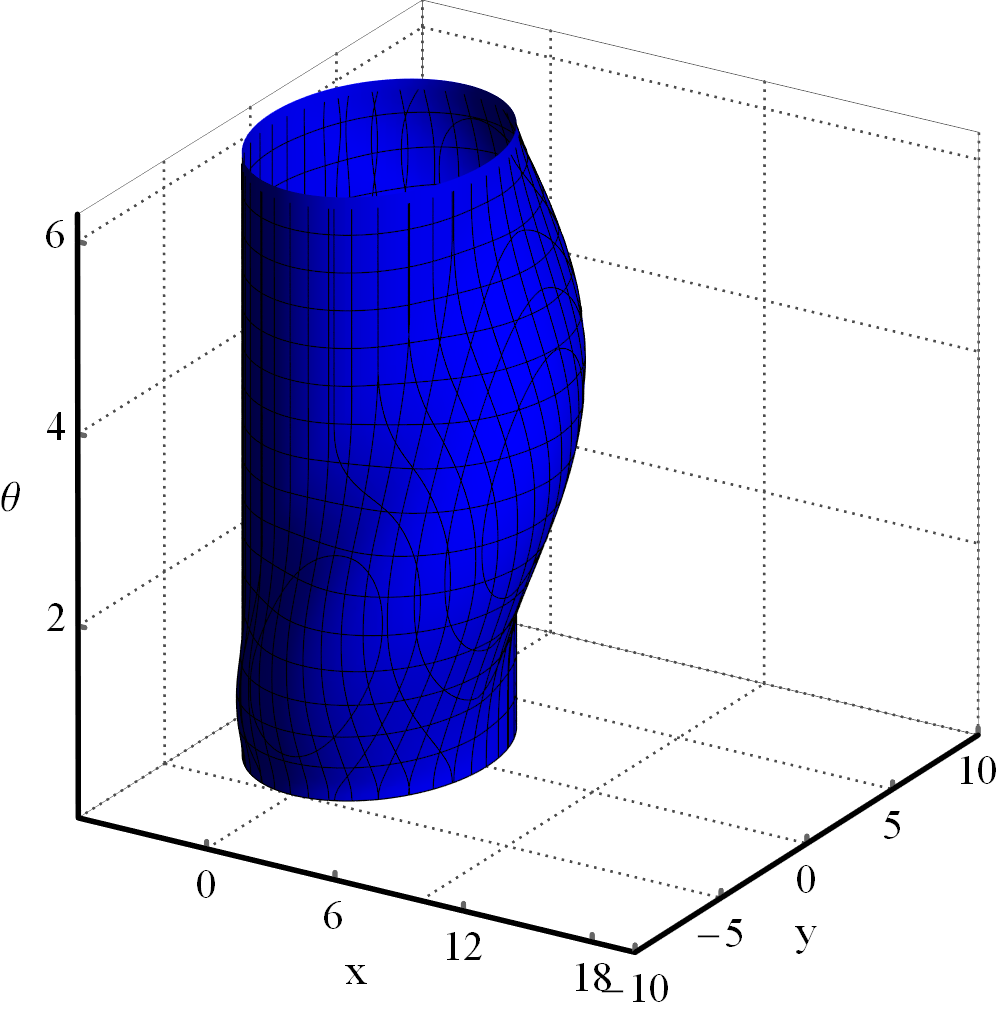}} 
		\subfigure[$R_K^c(J_2)$]{\includegraphics[width=0.24\textwidth]{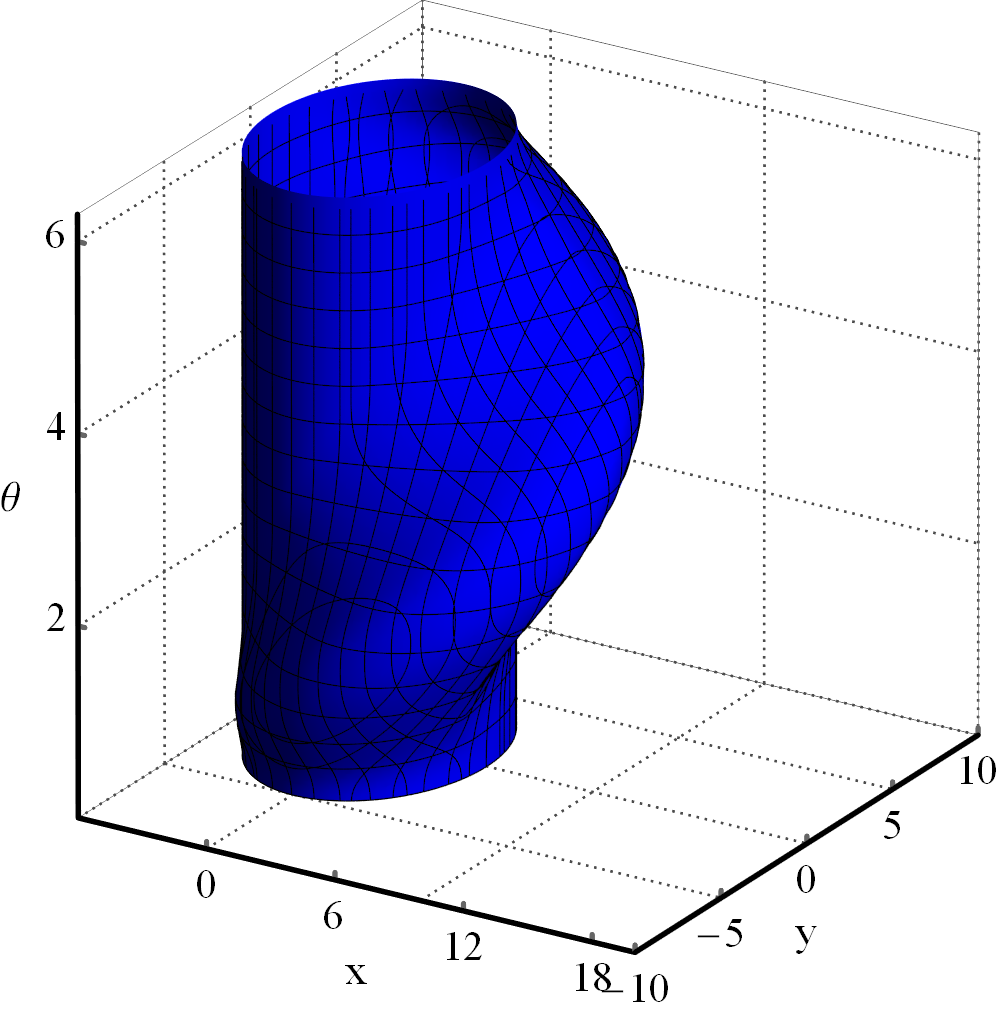}}\\
		\subfigure[$R_K^c(J_3)$]{\includegraphics[width=0.24\textwidth]{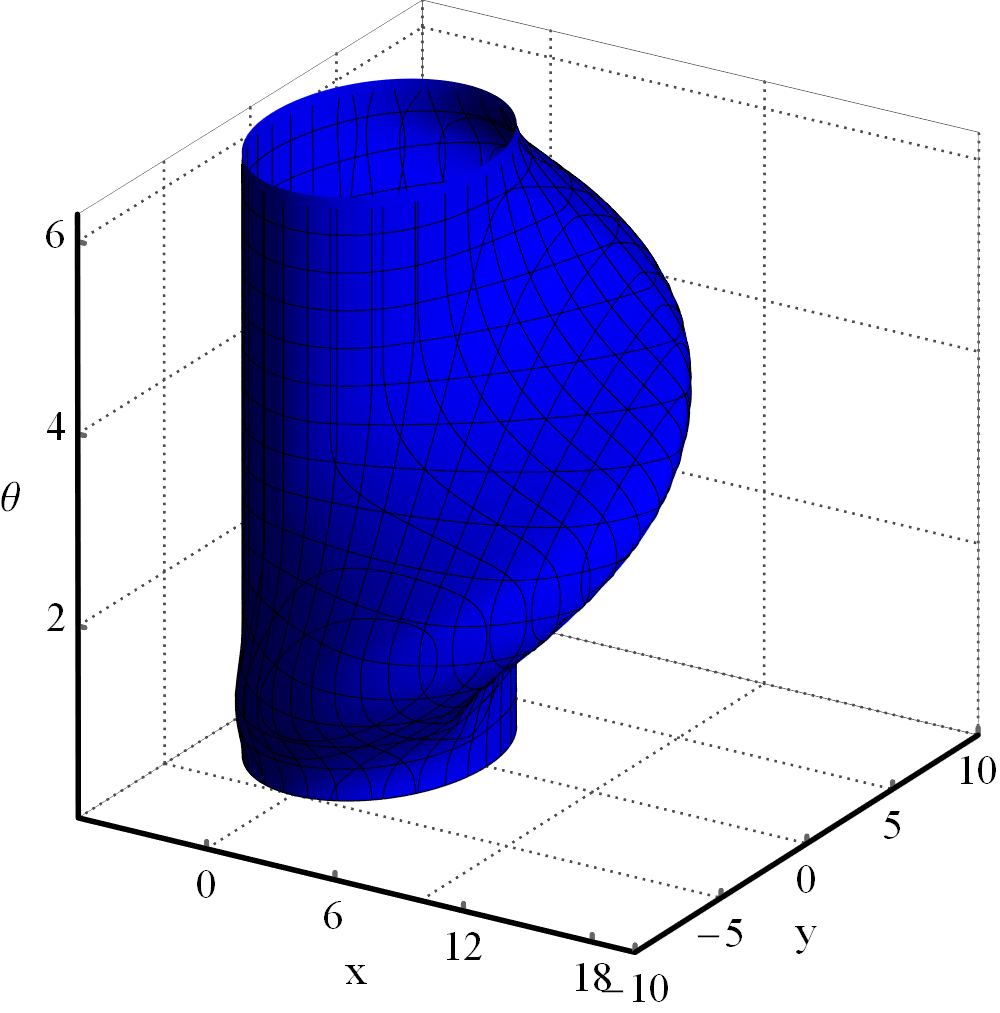}} 
		\subfigure[$R_K^c(J_4)$]{\includegraphics[width=0.24\textwidth]{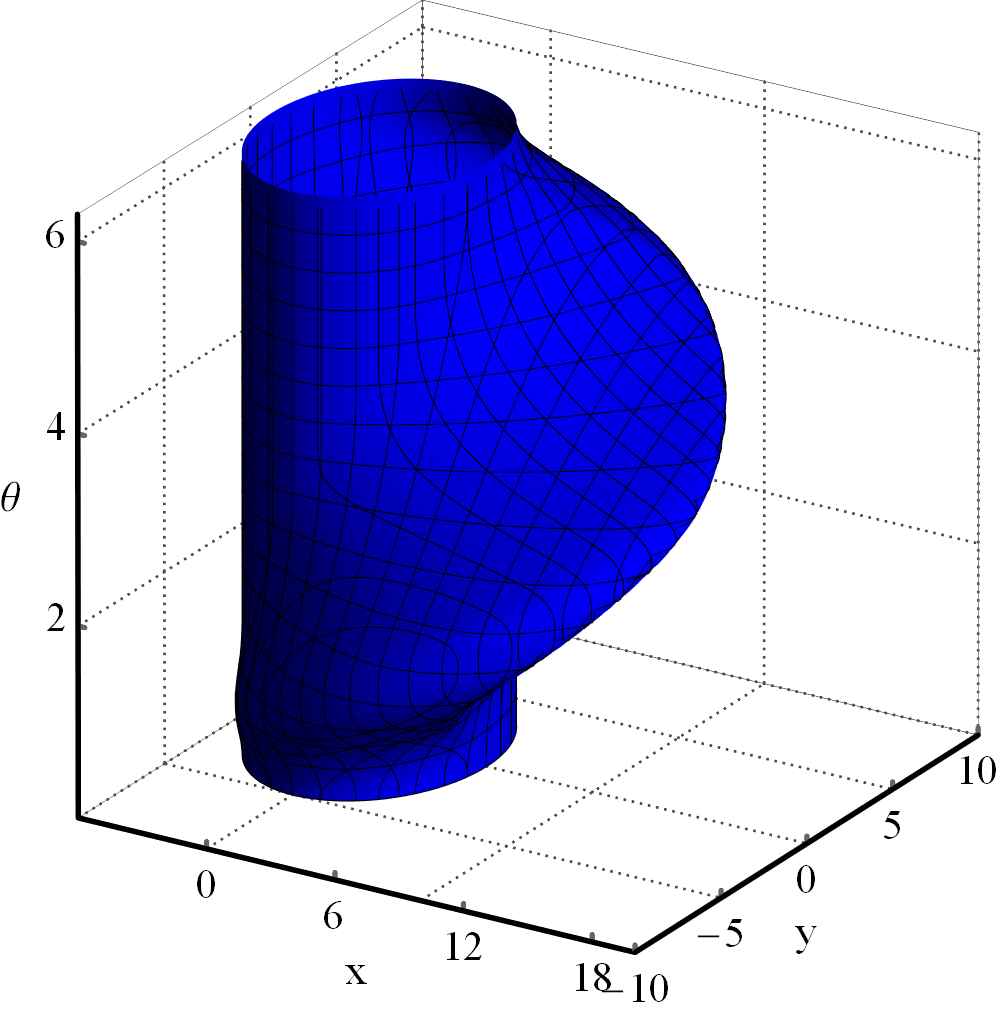}}
		\caption{Computation results of the CRTs with $\lambda=0.1$.}
		\label{figexp22}
	\end{figure}
	
	The latter case concerns the computation of the CRT. The solver settings of our method are the same as those in the previous case.
	The computation results of CRTs under $J_1,J_2,J_3$, and $J_4$ are shown in Fig. \ref{figexp22}.

	\section{Conclusions}
	\par This paper proposes a new method for computing the reachable tube for a two player, nonlinear differential games.
	In this method, a Hamilton-Jacobi equation with a running cost function is numerically solved, and 
	the reachable tube is described as a sublevel set of the viscosity solution of the Hamilton-Jacobi equation.
	
	An extension of the reachable tube 
	referred to as cost-limited reachable tube is newly introduced in this paper.
	A cost-limited reachable tube is a set of states that can be driven into the target set before the performance index grows to a given admissible cost. 
	Such a reachable tube can be obtained by specifying the corresponding running cost function
	for the Hamilton-Jacobi equation. 
	Another advantage of the proposed method is that it reduces the storage space consumption. 
	The cost-limited reachable tubes with different admissible costs (or the reachable tubes with different time horizons)
	can be characterized as a family of sublevel sets of the viscosity solution of the Hamilton-Jacobi equation at 
	some time point.
	
	The primary weakness of the proposed method for computing
	cost-limited reachable tube, and many other methods to compute
	reachable tube, is the exponential growth of memory and
	computational cost as the system dimension increases \cite{a008,i7}.
	Some approaches have been proposed to mitigate these costs, such as 
	projecting the reachable tube of a high-dimensional system into a collection of lower-dimensional 
	subspaces \cite{a002,i8,i9} and 
	exploiting the structure of the
	system using the principle of timescale separation and
	solving these problems in a sequential manner \cite{i10,i11,i12}.
	These will be considered
	in our future work.

% References

\bibliographystyle{Bibliography/IEEEtranTIE}
\bibliography{Bibliography/IEEEabrv,Bibliography/BIB_xx-TIE-xxxx}\ %IEEEabrv instead of IEEEfull

\end{document}